# Weak-link behavior of grain boundaries in superconducting Ba(Fe$_{1-x}$Co$_x$)$_2$As$_2$ bicrystals[*]

S. Lee[1], J. Jiang[2, a)], J. D. Weiss[2], C. M. Folkman[1], C. W. Bark[1], C. Tarantini[2], A. Xu[2], D. Abraimov[2], A. Polyanskii[2], C. T. Nelson[3], Y. Zhang[3], S. H. Baek[1], H. W. Jang[1], A. Yamamoto[2], F. Kametani[2], X. Q. Pan[3], E. E. Hellstrom[2], A. Gurevich[2], C. B. Eom[1], D. C. Larbalestier[2]

[1]*Department of Materials Science and Engineering, University of Wisconsin, Madison, WI 53706.*

[2]*Applied Superconductivity Center, National High Magnetic Field Laboratory, Florida State University, Tallahassee, FL 32310.*

[3]*Department of Materials Science and Engineering, University of Michigan, Ann Arbor, MI 48109.*

**Abstract:**

We show that despite the low anisotropy, strong vortex pinning and high irreversibility field $H_{irr}$ close to the upper critical field $H_{c2}$ of Ba(Fe$_{1-x}$Co$_x$)$_2$As$_2$, the critical current density $J_{gb}$ across [001] tilt grain boundaries (GBs) of thin film Ba(Fe$_{1-x}$Co$_x$)$_2$As$_2$ bicrystals is strongly depressed, similar to high-T$_c$ cuprates. Our results suggest that weak-linked GBs are characteristic of both cuprates and pnictides because of competing orders, low carrier density, and unconventional pairing symmetry.

PACS: 74.70.xa, 74.25.F-

___________________________________

The current-blocking effect of grain boundaries (GBs) with misorientation angles $\theta > 3\text{-}5^\circ$ has seriously delayed applications of high $T_c$ cuprates,[1-3] requiring the development of the biaxially textured $YBa_2Cu_3O_{7-x}$ (YBCO) conductors.[4,5] The pnictide superconductors, which encompass broad families of materials with critical temperatures $T_c$ up to 55K, $H_{c2}(0) > 100$ T, strong vortex pinning, moderate anisotropy $\eta = H_{c2}^{(c)}/H_{c2}^{(ab)} = 1\text{-}8$, and an irreversibility field $H_{irr}$ close to $H_{c2}$, are very promising for high-field applications,[6-10] leaving open one key question: can GBs transmit current? Here we report the explicit study of this vital property, using extensive transport, magneto-optical (MO), low-temperature scanning laser microscopy (LTLSM), and high resolution transmission electron microscopy (HRTEM) of $Ba(Fe_{1-x}Co_x)_2As_2$ (Ba-122) epitaxial thin-film bicrystals. We show a strong intrinsic current-blocking effect of GBs, across which the critical current density $J_b(\theta)$ rapidly decreases as the misorientation angle $\theta$ between neighboring grains increases.

We grew epitaxial ~350 nm thick Ba-122 thin films in-situ on [001] tilt (100) $SrTiO_3$ bicrystal substrates using pulsed laser deposition (PLD) with a KrF (248 nm) UV excimer laser in a vacuum of 2 μTorr. The PLD target was prepared by solid state reaction with nominal composition of Ba:Fe:Co:As = 1 : 1.84 : 0.16 : 2.2. Four-circle x-ray diffraction showed excellent epitaxy with cube-on cube, in-plane epitaxial relationship with the substrates.[11] The misorientation angles in the Ba-122 film were determined by off-axis azimuthal θ-scans of the 112 reflections and found to be 2.5°, 5.7°, 8.5° and 24° (hereafter approximated to 3°, 6°, 9° and 24°), identical to the $SrTiO_3$ bicrystal substrates.

Epitaxial pnictide films have so far been hard to produce, especially the F and O doped highest $T_c$ RE1111 phase. Understanding GB properties in polycrystals has been compromised by their multi-phase nature and complex microstructure. Significant intergranular transport across randomly misoriented grains of RE1111 bulks was deduced, but almost all GBs were coated with metallic FeAs phases, which mask the intrinsic GB properties.[12-16] Hiramatsu et al. reported epitaxial Co-doped Sr122 thin films grown on (001) (La,Sr)(Al,Ta)$O_3$ substrates but no $J_c$ measurements.[17] Figure 1 shows TEM images of a 24° bicrystal. The film is uniform in thickness over a large area, while the HRTEM image shows the (001) planes are well aligned across the GB.

Figure 2 shows intergranular resistance curves $R(H,T)$ across a 6° [001] tilt GB in fields up to 35 T. The resistive transitions are sharp, confirming the high quality of our bicrystal films. Moreover, the $R(T)$ curves remain narrow under strong fields indicating a low-$T_c$ depinning resistive transition with insignificant vortex fluctuations, unlike the thermally-activated vortex dynamics in cuprates and Nd-1111 pnictides.[7] By contrast, Figure 2 shows that $H_{irr}(T)$ defined by $R(T,H_{irr}) = 0.1R_n(T_c)$ is close to $H_{c2}$ defined by $R(T,H_{c2}) = 0.9R_n(T_c)$, as in previous studies of 122 single crystals.[8-10] Here $R_n$ is the normal state resistance. $H_{c2}(T)$ and $H_{irr}(T)$ are very similar for both grain and GB links.

Shown in Figure 3 are representative LTLSM and MO images of 6° and 9° bicrystals, which demonstrate the significant current-blocking effect of even low-angle GBs. The LTLSM image of the electric field E of the transport current flowing through GB in Fig. 3a shows a 23 times enhancement of E at the 6° [001] tilt GB. The MO image in Fig. 3b shows that the 9° [001] tilt GB is also a strong obstacle for the magnetization currents, which make a nearly 90° turn at the GB because it can transmit only about 10% of the intragrain critical current.

We performed detailed studies of the grain and GB critical current densities $J_c(T,B)$ and $J_{gb}(T,B)$ for all the bicrystals in fields up to 16 T. Figure 4 shows that $J_{gb}(12K, 0.5T)$ falls off by an order of magnitude as θ increases from 3 to 24°. This qualitative behavior is similar to $J_{gb}(\theta)$ for [001] tilt GBs in YBCO, as shown in the inset. Similar to YBCO, the 3° GB in $Ba(Fe_{1-x}Co_x)_2As_2$ does not obstruct supercurrent, while at higher angles $J_{gb}(\theta)$ becomes much smaller than the grain $J_c$. We do not yet have enough data in the crucial region of $9° < \theta < 30°$ to conclude whether $J_{gb}(\theta)$ drops exponentially as it does in cuprates,[1] but the fact that even low-angle GBs with $\theta > 6°$ do obstruct supercurrent in pnictides is quite clear.

The field dependence of $J_{gb}(T,H)$ shown in Figure 5 exhibits characteristics typical of cuprate GB transport. At low fields, $J_{gb}(H)$ is smaller than the grain $J_c(H)$, but the difference between them diminishes as $H$ increases, tending to make GBs lesser obstacles at higher fields. This behavior, well established for low-angle YBCO bicrystals is due to pinning of Abrikosov-Josephson vortices on GBs and their interaction with vortices in the grains.[18-20]

The totality of our data unambiguously indicates that GBs in 122 pnictides exhibit current-limiting behavior similar to that observed in high-$T_c$ cuprates. The relative suppression of superconductivity at GBs is determined by the ratio, $J_{gb}(t)/J_d(t)$ at the reduced temperature $t = T/T_c$ where $J_d(T) = \phi_0/3^{3/2}\mu_0\lambda^2\xi$ is the depairing current density, $\phi_0$ is the magnetic flux quantum, $\lambda$ is the London penetration depth, and $\xi$ is the in-plane coherence length.[21] YBCO and Ba-122 have similar zero-temperature values of $\lambda(0) \approx 150$-$200$ nm and $\xi(0) \approx 2$ nm.[7,10] Taking, for example, $J_{gb} \approx 10^8$ A/m$^2$ for a 24° [001] tilt GB in YBCO at 77K (t = 0.84) and $J_d(77K) \approx 4\times10^{11}$ A/m$^2$,[1] we obtain $J_{gb}(t)/J_d(t) \approx 2.25\times10^{-4}$. In turn, for a 24° [001] tilt GB in Ba-122 at 16K (t = 0.76), our data in Figures 3 and 4 give $J_{gb} \approx 10^8$ A/m$^2$, and $J_d(16K) \approx 2\times10^{11}$ A/m$^2$ (for $\lambda(t) \approx 160(1-t)^{-1/2} = 330$ nm and $\xi(t) \approx 2.44(1-t)^{-1/2} = 4.6$ nm).[10] Thus, the ratio $J_{gb}(t)/J_d(t) \approx 5\times10^{-4}$ for Ba(Fe$_{1-x}$Co$_x$)$_2$As$_2$ turns out to be rather small and of the same order of magnitude as that for YBCO, suggesting that high angle GBs in pnictides can be regarded as Josephson weak links.

We now briefly discuss possible mechanisms underlying our results. Both for the d-wave cuprates and for multiband pnictides, GB crystalline disorder depresses superconducting properties because of the short coherence lengths, which result from their small Fermi velocities and low carrier densities. The values of $\xi(0) \approx 1$-$2$ nm extracted from $H_{c2}$ measurements are indeed similar for cuprates and pnictides despite significant $T_c$ differences.[7-10] Depression of multiband superconductivity at GBs in pnictides could result from strong interband scattering by imperfect GB structural units in the chain of GB dislocation cores or by strain- or charge-driven Cottrell atmospheres of segregated impurities at the GB.[22,23] Strong antiferromagnetic correlations in the Fe-As planes and GB impurity segregations may also induce local magnetic nanostructures, which would also contribute to the weak link behavior of GBs in pnictides. Given the strong coupling of the superconducting order parameters in different bands,[24] pairbreaking magnetic and nonmagnetic scattering at the GB may play a much greater role in the order parameter suppression at GBs in 122 than in the two-band superconductor $MgB_2$ in which weak interband coupling and interband impurity scattering are consistent with the lack of weak link behavior of GBs. Current blocking GBs in the pnictides may also result from strong shear strains near GB dislocation cores, which distort the local angle between Fe-As bonds from its optimal value at which $T_c$ is maximum.[25,26] Any or all of these effects may suppress superconductivity in the channels between dislocations in low-angle GBs. Another common mechanism results from the generic phase diagram of cuprates and pnictides, which contains the ubiquitous superconducting 'dome" around optimal doping. As in the cuprates, local compositional variations, strain and charging effects on the GB may depress the order parameter as the superconducting state moves away from optimal doping towards the competing parent antiferromagnetic phase. Also charging effects at dislocation cores, which control superconductivity suppression at GBs, are greatly amplified by the long Thomas-Fermi screening length $\approx \xi$ in the low carrier density pnictides and cuprates.[21]

Our results suggest that understanding the potential for local overdoping of GBs to ameliorate GB current blocking effects in pnictides will likely be as important as in cuprates. Since the qualitative mechanisms of GB weak-link behavior discussed above may be insensitive to details of their pairing scenarios and superconducting order parameter symmetry, conditions favorable for weak-linked GBs may be present in any high-$T_c$ superconductor with competing orders, short coherence length and low carrier density.

We are grateful to John Fournelle, Marina Putti, and Pei Li for discussions and experimental help. Work at UW was supported by DOE grant DE-FG02-06ER463. Work at NHMFL was supported under NSF Cooperative Agreement DMR-0084173, by the State of Florida, and by AFOSR grant FA9550-06-1-0474. EM work at the University of Michigan was supported by the Department of Energy under grant DE-FG02-07ER46416. AY was supported by a fellowship of the Japan Society for the Promotion of Science.

**Figure Captions**

FIG 1. TEM micrograph of a cross section of the GB region in a 24° [001] tilt Ba-122 film which confirms that the grains are epitaxial and c-axis oriented. The HRTEM image (a) is viewed parallel to the GB in longitudinal section parallel to the (001) planes. The (001) planes are well aligned across

the GB. A lower magnification image from a region with a single grain orientation (b) shows that the film is of uniform thickness. The selected area diffraction pattern of the Ba-122 film along the [010] zone axis is shown in (c) together with the corresponding SAD pattern for the SrTiO3 substrate (d). The SAD patterns show that the Ba-122 film is single-crystal, c-axis oriented, and epitaxially related to the substrate.

FIG. 2. Irreversibility field $H_{irr}$ and upper critical field $H_{c2}$ defined by the 10% and 90% points of $R_n(T_c)$ on resistance curves $R(T,H)$ across a 6° [001] tilt GB and its neighboring grain with field applied perpendicular to the film. Inset: resistive transitions up to 35 T across the GB.

FIG. 3. (a) LTLSM image scanned in 1 μm steps of a 6° [001] tilt bicrystal taken at 12 K and 0.25 T with bias current of 39 mA which generated 17 μV across the bridge. It shows a 2D map of $E(x,y)$ induced by transport current along the 100 μm wide x 700 μm long bridge across the GB. E is enhanced by a factor of 23 at the GB as compared to the intragranular E. (b) MO image of a 9° bicrystal with computed streamlines (yellow) of the magnetization current calculated from the Bean model [Ref. 27]. The MO image was taken after cooling to 6 K in a perpendicular field of 120 mT and then reducing the field to zero.

FIG. 4. Dependence of the critical current density across GBs $J_{gb}$ (12 K, 0.5 T) as a function of the misorientation angle θ. The inset shows summary data from Ref. 1 for YBCO GBs.

FIG. 5. Comparative field dependencies of the critical current density across GB $J_{gb}(H)$ and $J_c(H)$ in the grains for (a) 6° and (b) 24° bicrystals at 12K, 14K and 16K. The gap between $J_{gb}(H)$ and $J_c(H)$ diminishes as the field increases.

**Figure 1**

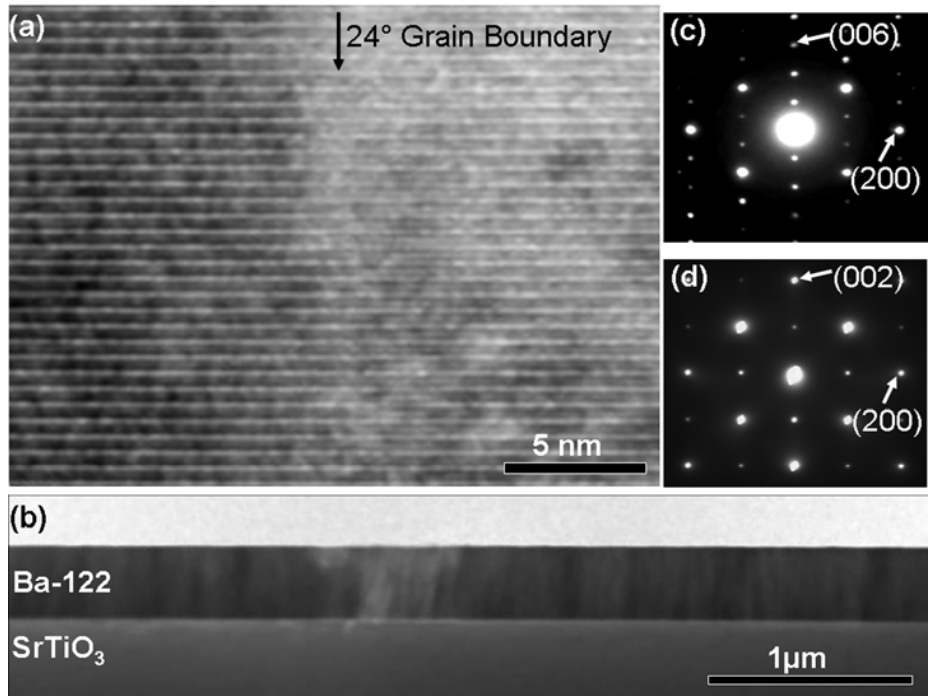

**Figure 2**

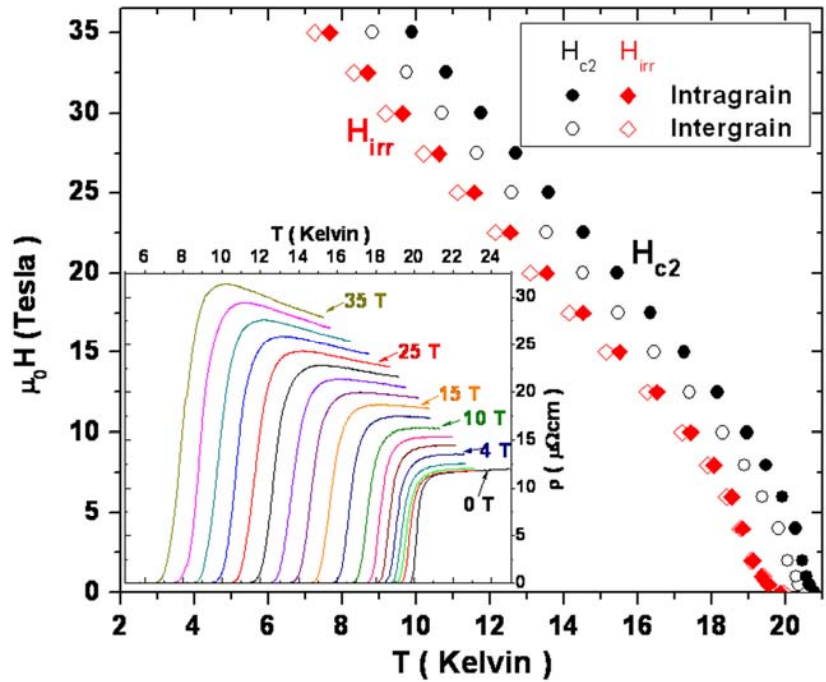

**Figure 3**

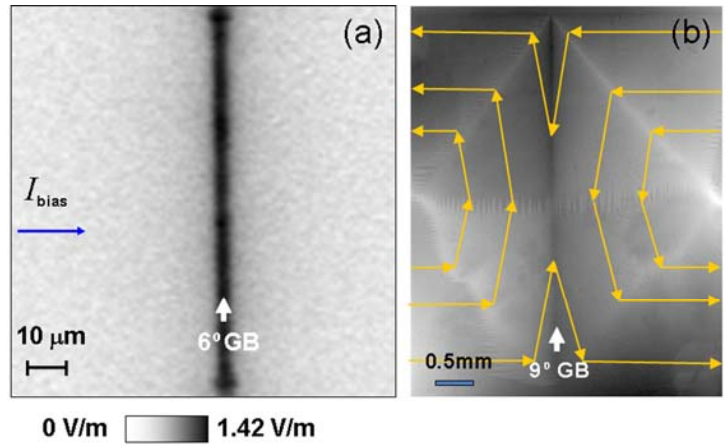

**Figure 4**

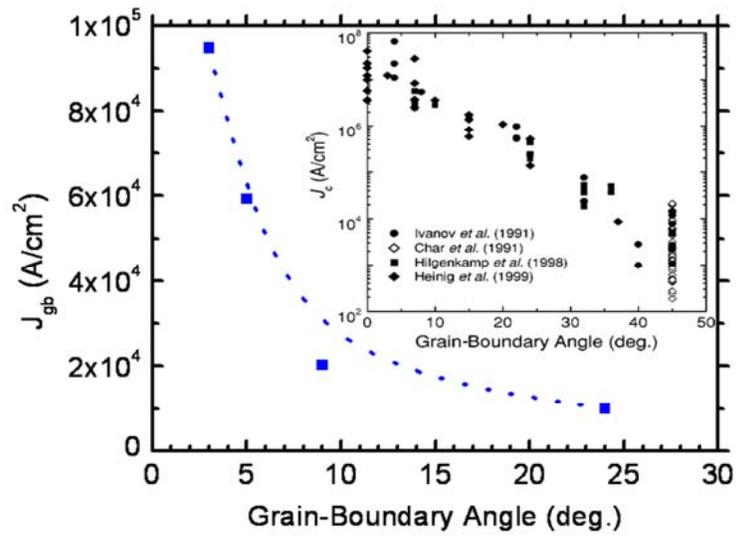

**Figure 5**

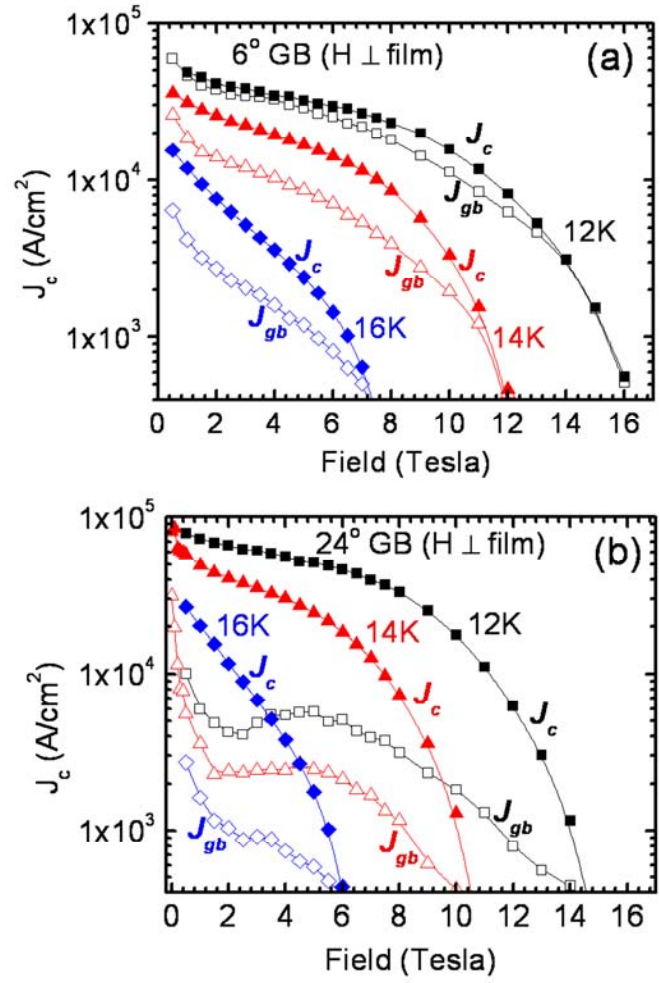